\documentclass[%
reprint,
amsmath,amssymb,
aps, 
superscriptaddress,
longbibliography, 
floatfix,
textcomp,mathcomp
]{revtex4-2}

\usepackage[dvipsnames]{xcolor}
\usepackage{dcolumn}
\usepackage{bm}
\usepackage{natbib}
\usepackage{makecell}
\usepackage{svg}
\usepackage{float}
\usepackage{multirow, tabularray}
\usepackage[caption=false]{subfig}
\usepackage[normalem]{ulem}
\useunder{\uline}{\ul}{}

\usepackage{amsmath,amssymb,amsfonts}
\usepackage{algorithmic}
\usepackage{graphicx,caption}
\usepackage{textcomp}
\usepackage{xcolor}

\usepackage{pgf}
\usepackage{tikz,tikzscale}
\usetikzlibrary{arrows,positioning,calc,shapes,arrows.meta,external}
\usepackage{pgfplots}
\pgfplotsset{compat=newest}
\usepgfplotslibrary{statistics,fillbetween}
\definecolor{tab_blue}{RGB}{31,119,180}
\definecolor{tab_orange}{RGB}{255,127,14}
\definecolor{tab_green}{RGB}{44,160,44}
\definecolor{tab_red}{RGB}{214,39,40}
\definecolor{tab_purple}{RGB}{148,103,189}
\definecolor{tab_brown}{RGB}{140,86,75}
\definecolor{tab_pink}{RGB}{227,119,194}
\definecolor{tab_gray}{RGB}{127,127,127}
\definecolor{tab_olive}{RGB}{188,189,34}
\definecolor{tab_cyan}{RGB}{23,190,207}

\graphicspath{{./images/}}

\newcommand{\mics}{\,$\mu$s\:}

\newcommand{\affilUNM}[0]{\affiliation{Dept. of Electrical and Computer Engineering, The University of New Mexico, Albuquerque, NM, USA}}
\newcommand{\affilSandia}[0]{\affiliation{Sandia National Laboratories, Albuquerque, New Mexico 87185, USA}}

\def\BibTeX{{\rm B\kern-.05em{\sc i\kern-.025em b}\kern-.08em
    T\kern-.1667em\lower.7ex\hbox{E}\kern-.125emX}}
    
\begin{document}
\title{Scatter-Gather DMA Performance Analysis within an SoC-based Control System for Trapped-Ion Quantum Computing}

\author{Tiamike Dudley}\email{tiadud@unm.edu}\affilUNM
\author{Jim Plusquellic}\affilUNM
\author{Eirini Eleni Tsiropoulou}\affilUNM
\author{Joshua Goldberg}\affilSandia
\author{Daniel Stick}\affilSandia
\author{Daniel Lobser}\affilSandia

\begin{abstract}
Scatter-gather dynamic-memory-access (SG-DMA) is utilized in applications that require high bandwidth and low latency data transfers between memory and peripherals, where data blocks, described using buffer descriptors (BDs), are distributed throughout the memory system. The data transfer organization and requirements of a Trapped-Ion Quantum Computer (TIQC) possess characteristics similar to those targeted by SG-DMA. In particular, the ion qubits in a TIQC are manipulated by applying control sequences consisting primarily of modulated laser pulses.  These optical pulses are defined by parameters that are (re)configured by the electrical control system. Variations in the operating environment and equipment make it necessary to create and run a wide range of control sequence permutations, which can be well represented as BD regions distributed across the main memory. In this paper, we experimentally evaluate the latency and throughput of SG-DMA on Xilinx radiofrequency SoC (RFSoC) devices under a variety of BD and payload sizes as a means of determining the benefits and limitations of an RFSoC system architecture for TIQC applications.
\end{abstract}

\maketitle

\section{Introduction}
\label{sec:introduction}
A trapped-ion quantum computer (TIQC) \cite{clark:2021} uses modulated laser pulses to precisely control the quantum state of its trapped ion qubits. The sequence of laser pulses applied to an ion is referred to as a gate sequence, where gates represent the logic operations that together form a quantum algorithm. The magnitude, phase, and frequency of the laser pulses are commonly controlled using optical modulators with radiofrequency (RF) signal inputs, which can be generated using benchtop instruments like arbitrary waveform generators (AWGs) or direct-digital synthesizers (DDS), or using high-speed digital-to-analog converters (DACs) embedded within a device like a radiofrequency System-on-Chip (RFSoC) device.
The programmable logic (PL) component of the RFSoC is capable of emulating AWG and DDS functionality, while providing a highly diverse mechanism to configure gate sequence pre-processing pipelines. 

An important constraint to fully leveraging the flexibility of the PL for TIQC applications is ensuring the sufficient speed of data transfers between components of the memory hierarchy within the RFSoC system architecture. This was measured at a general level for the RPU to PL in \cite{irtija:2023}, but here we expand that by analyzing communication in the particular context of scatter-gather dynamic-memory-access (SG-DMA). As described in this paper, SG-DMA provides a path to making practical TIQC control systems that have the flexibility to define pulses on the fly with latencies that do not cause significant delays in quantum algorithms, most importantly quantum error correction.

The memory hierarchy associated with the Xilinx RFSoC ZCU111 used in this work is layered in a configuration commonly used in high-end microprocessors, with two 4 GB DDR4 (DRAM) defining the largest, and lowest layer of the hierarchy, and 2-level processor caches representing the smaller and fastest layers. However, SoC architectures augment the memory hierarchy by adding PL-side block RAM (BRAM/UltraRAM), lookup-table (LUT) RAM, and distributed RAM, represented as flip-flops in the PL fabric. 

The large size of the DRAM components can be leveraged to accommodate a wide range of gate sequence definitions, which are configured in the proposed architecture by application processing units (APUs). A real-time processing unit (RPU) is charged with transferring APU-pre-configured gate sequences to the PL-side memory components by issuing commands to a dynamic memory access (DMA) engine, configured in scatter-gather (SG) mode. Scatter-gather mode utilizes a set of buffer descriptors (BDs), arranged in a BD ring, with each pointing to an arbitrary memory region within the DRAM. The primary focus of this paper is on evaluating SG-DMA performance characteristics, i.e., latency and throughput, as a function of several BD ring parameters, including the number of BDs, their buffer size, and their location in DDR.

The RF signals used to define quantum gates in a TIQC are derived directly from the gate sequence data. Therefore, SG-DMA performance is the limiting factor on meeting the timing requirements associated with synchronization and phase coherence, and on implementing feedback-based calibration within the TIQC system. An important overall goal of our system is to arbitrarily reconfigure gate parameters based on measurements made on feedback channels within a time period that is on the order of the shortest gate time, assumed here to be lower-bounded by $\approx$1\mics for a single qubit gate. The inside loop of the control system consists of 
state preparation followed by a sequence $i$ of control pulses, $P_{i}(X_{j})$, with $X_{j}$ defining a set of pulse parameters. Although adjustments to the pulse parameters occur at longer time scales within a calibration loop, a high-speed data streaming interface is required to meet data delivery requirements associated with the inside loop.

The latency and throughput requirements for the memory-mapped DRAM to PL streaming interface implemented by the SG-DMA engine depends on several factors, including the size of the $X_{j}$ gate sequence data packets, the duration and complexity of the laser pulses, the number of channels and corresponding qubits serviced by the control system, the periodic stalls needed to update pulse parameters based on calibration measurements, and the frequency at which intra-algorithm measurements affect future gates. Although it is necessary for the SG-DMA engine to accommodate continuously changing gate sequences, it is common that a particular sequence is repeated multiple times before an update to the pulse parameters is necessary. The proposed control system incorporates several optimizations that reduce bandwidth in these cases, as described in the following sections.

The following characterizes the main contributions of this work. The relationship between the measured performance characteristics and TIQC system requirements is interleaved with the presentation of the results.
\begin{itemize}
 \item A multi-processor system architecture which utilizes APU, RPU, and PL is used as the base architecture for the performance analysis. Performance trade-offs associated with using different architectural components for SG-DMA setup and operation are discussed, along with their respective strengths and weaknesses. 
 \item A latency and throughput analysis and statistical characterization is carried out over an extensive range of SG-DMA parameters, including BD ring size, BD block size, DMA bus-width, etc. Architectural features impacting performance as they relate to SG-DMA parameters are discussed. 
\end{itemize}

\section{Related Work}
\label{sec:background}

Other research has investigated the use of SG-DMA for systems with similar demands of the rapid transfer of different amounts and regions of data.

The system described in \cite{Fjeldtvedt:2019} used a customized DMA engine to transfer three-dimensional blocks of data for hyperspectral imaging applications. The CubeDMA system assumed a contiguous cube-shaped memory region and can compute the addresses of data in flight, while SG-DMA requires a separate data structure to obtain the addresses of data. Thus SG-DMA can move discontiguous regions of memory and may suffer a performance penalty as a result.

To measure and mitigate memory interference, the systems in \cite{Brilli:2022} and \cite{Alismail:2023} ran PS and PL benchmarks in parallel to measure and improve slowdowns due to memory contention and inefficiencies to access DDR. The former used Controlled Memory Request Injection (CMRI) to increase the utilization of the DDR. The latter system used scheduling policies within the PL to reduce memory stall times.

The work in \cite{vaishnav:2019} presented an analysis of the throughput and latency of AXI port configurations for the ZCU102 and Ultra-96 boards. Their analysis considered AXI bus width, burst size, memory chip configuration, access patterns, and transaction frequency. An analysis of DMA engine performance using the ZCU102 is presented in \cite{argyriou:2021}. Their work observed how the DMA parameters of burst and memory stride sizes affected the performance of the DMA engine.

In our work, we utilize the ZCU111 for performance characterization of SG-DMA operating in memory-mapped-to-streaming (MM2S) mode, and a multi-processor system architecture consisting of an APU, RPU, and the PL. We report the impact of using different SG-DMA parameters on performance, as well investigate inter-processor communication benefits and limitations using general purpose input-output (GPIO) and the RPMsg API.

\section{Communication}
\label{sec:communication}

\begin{figure}[t]
   \centering
    \includegraphics[width=0.5\textwidth]{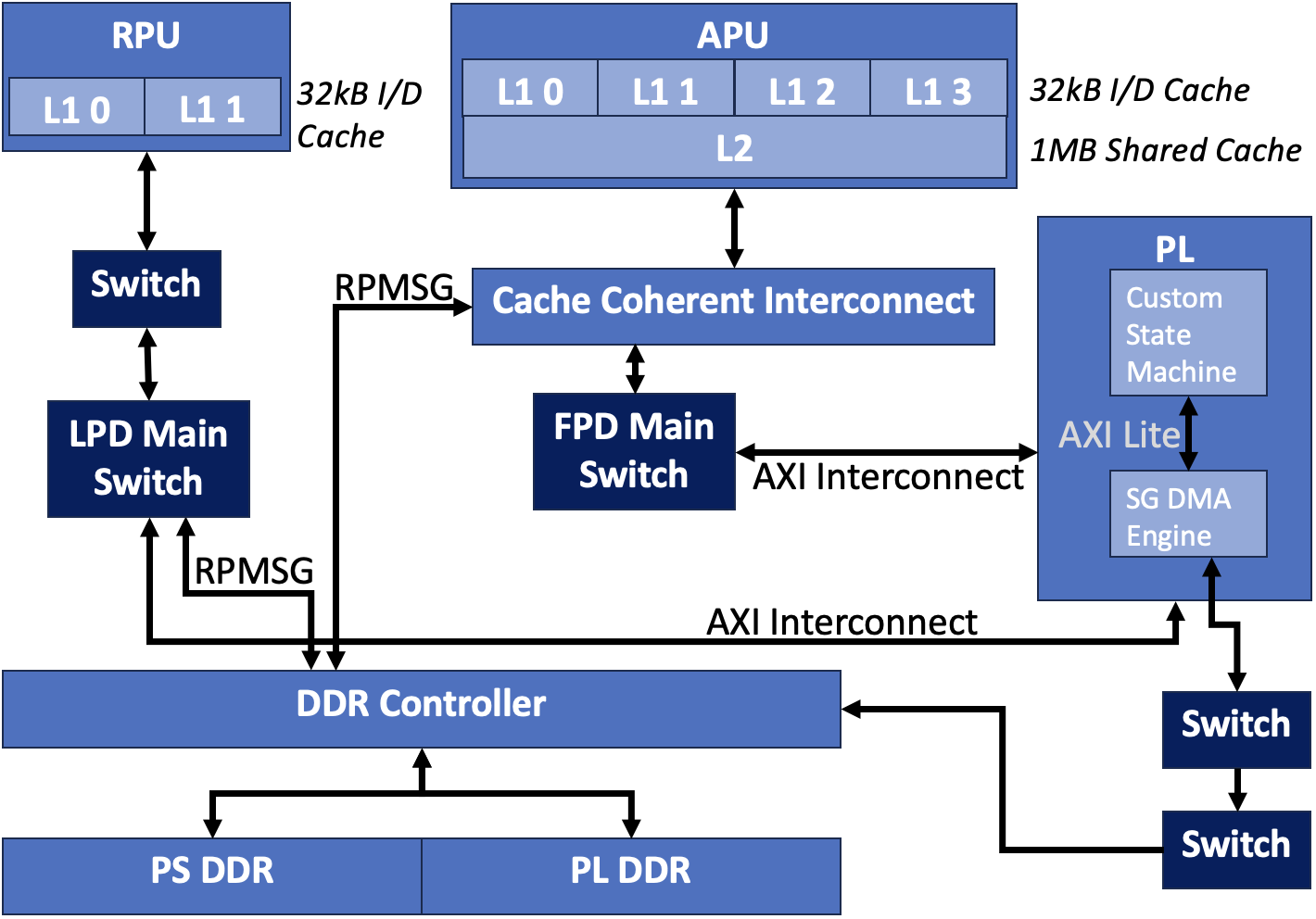}
   \caption{Zynq MPSOC system architecture showing microprocessors, programmable logic, DDR, and AXI channels used.}
   \label{fig:ZCU111_Architecture}
\end{figure}

\subsection{System Architecture} \label{sec:SystemArch}

A block diagram showing the processing and interconnect components within the Zynq UltraScale+ RFSoC on the ZCU111 board is shown in Fig. \ref{fig:ZCU111_Architecture}. The SoC consists of a dual-core Cortex R5 RPU, a quad-core Arm Cortex A53 APU, a PL region, and two external DDR4 DRAM memories, interconnected through a complex point-to-point network of Arm Advanced eXtensible Interface (AXI) switches. The DDR memories are accessible to both RPU and APU for system instantiations in which the memory map for both DRAMs are configured with overlapping address spaces. Otherwise, the PL DRAM is not accessible by the 32-bit RPU when address-mapped above the 4 GB PS DRAM. Each core in the RPU possesses a 32 kB L1 instruction and data cache, while each core in the APU possesses a 32 kB L1 instruction and data cache and a shared 1 MB L2 cache.

The Vivado block diagram tool is used to create a system level diagram which utilizes the APU, RPU, block RAM (BRAM), DMA, GPIO and PL DDR. The DDR is created using the memory interface generator (MIG), and is interconnected through AXI to other components of the system architecture. In our experiments, we map a portion of the PL DRAM into the address space of the APU and RPU at an aperture given by 0x80000000 to 0x9FFFFFFF. This makes the PS and PL DDR accessible by both the APU and RPU. An AXI Smart-connect block is used to access the PL DDR, which is capable of arbitrating between multiple master IP interfaces.

The DMA engine possesses several input-output interfaces. The configuration and control channel of the DMA engine is connected to an AXI-lite interconnect and is controlled by the RPU in our experiments. The memory-mapped interfaces of the DMA engine connect to the PL DDR while the streaming interfaces are connected to a custom state machine (SM) (described below). The SM stores incoming data from the MM2S slave interface to a BRAM, and writes BRAM out-going data to the streaming-to-memory-mapped (S2MM) master interface. Two GPIO IP blocks are instantiated and are accessible by the APU, RPU, and SM. Although not shown in Fig. \ref{fig:ZCU111_Architecture}, a Linux kernel is created that enables the RPU and APU to communicate using an API defined by the libmetal and OpenAMP (\textit{asymmetric} multiprocessing) standards \cite{OpenAMP} called RPMsg. RPMsg utilizes on-chip tightly-coupled memory for code, stack, heap, etc., and the PL DDR for inter-processor communication.

\subsection{Scatter Gather DMA (SG-DMA)} \label{sec:ScatterGatherDMA}

The DMA IP block provided by Xilinx \cite{SG-DMA} is implemented entirely in the PL, and can be configured to operate in either simple mode or scatter gather (SG) mode. In previous work, we investigated the latency and throughput of simple mode across a range of data transfer widths and PL clock frequencies \cite{irtija:2023}. In this work, we present a statistical characterization of latency and throughput with DMA configured in SG mode.

\begin{figure}
   \centering
   \includegraphics[width=0.3\textwidth,keepaspectratio=true]{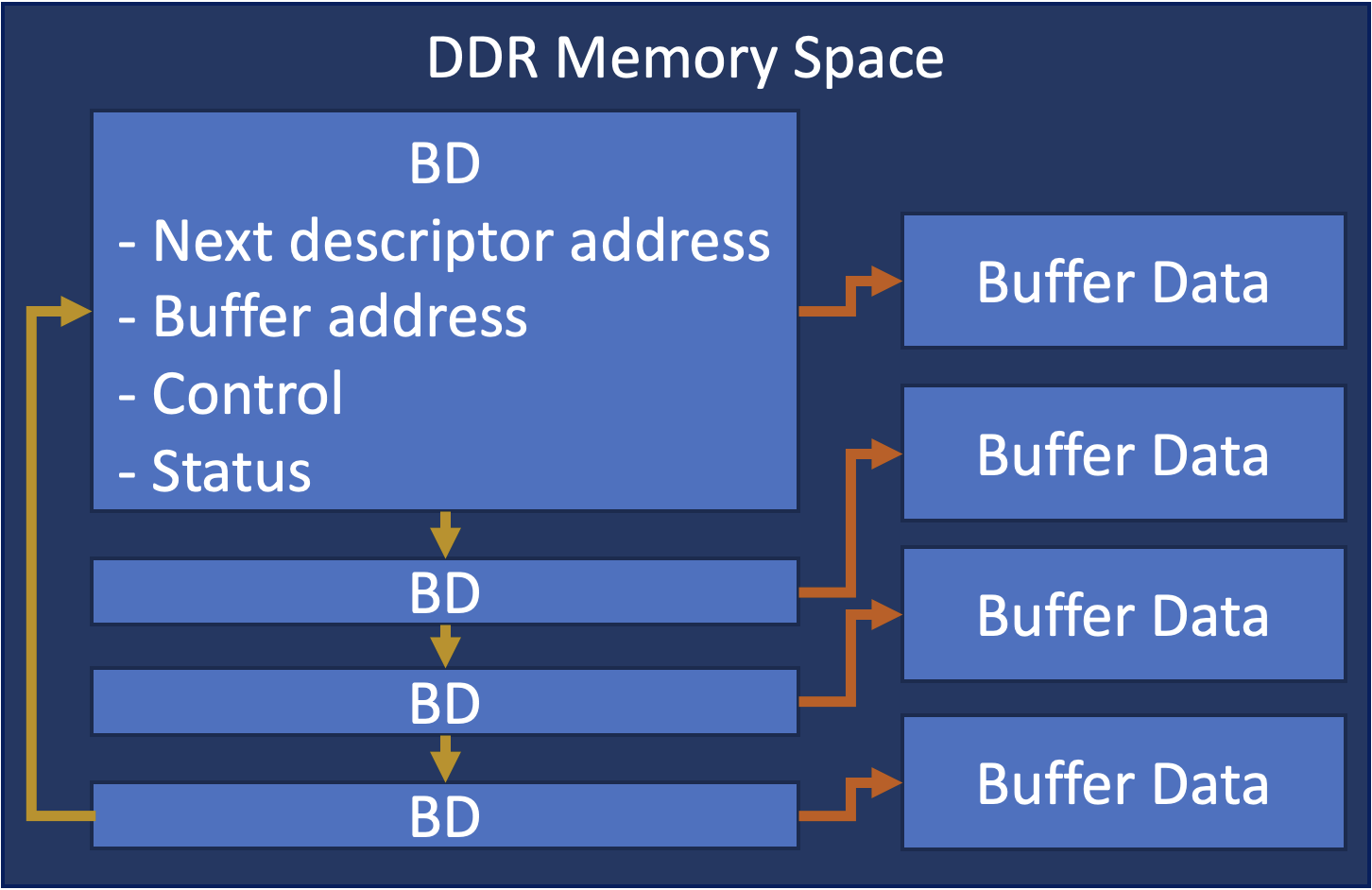}
   \caption{Structure of BD ring in DDR.}
   \label{BD-Ring_data_structure}
\end{figure}

Unlike simple mode, SG mode allows data blocks to be distributed across the DDR at non-sequential addresses. Data blocks are described using BDs, which are stored in a ring as a linked-list data structure. Each BD is 64 bytes in size and contains a 64-bit base address, a 26-bit size field (for up to 64 MB payload sizes), a pointer to the next BD and several control and status fields. An example of a BD ring is shown in Fig. \ref{BD-Ring_data_structure}. 

The proposed TIQC system uses the APU to construct gate sequences in PS and/or PL DDR. The linked list data structure associated with the BD ring provides a fast and flexible mechanism for creating heterogeneous gate sequence orderings in real time during system operation, that can be tuned based on feedback from instrument sensors. After the creation of each custom sequence by the APU, the base address of the BD ring is transferred to the RPU, using either RPMsg or GPIO. The RPU is tasked with configuring the SG-DMA engine with the base address and then starting the SG-DMA engine, which occurs when the tail descriptor is transferred to the SG-DMA engine. Given the tight timing requirements of TIQC systems, we investigate the latency and throughput characteristics of this architecture using a wide range of BD ring and SG-DMA configurations. 

\begin{table*}[htb]
\centering
\begin{tabular}{|c||c|c||c|c||c|c||c|c|}
\hline
     Data Width (b) & LUTs Utilized & \% of total & \makecell{LUTRAM\\Utilized} & \% of total & FFs Utilized & \% of total & \makecell{BRAM Pages\\Utilized} & \% of total \\
\hline
\hline
     32 & 10489 & 2.47 & 1970 & 0.92 & 15523 & 1.83 & 3 (90kb) & 0.28\\
\hline
     64 & 10639 & 2.50 & 1967 & 0.92 & 15792 & 1.86 & 5 (162kb) & 0.46\\
\hline
     256 & 13598 & 3.20 & 2695 & 1.26 & 20033 & 2.36 & 9 (306kb) & 0.83\\
\hline
     512 & 16120 & 3.79 & 4158 & 1.95 & 25420 & 2.99 & 9 (306kb) & 0.83\\
\hline
     1024 & 22864 & 5.38 & 7109 & 3.33 & 35543 & 4.18 & 9 (306kb) & 0.83\\
\hline
\end{tabular}
\caption{Utilization of a dual-channel SG DMA on the ZCU111 RFSoC as a function of data bus width. A minimal functional implementation requires an instance of the MPSoC, AXI Interconnect, AXI SmartConnect and the Processor System Reset IP block.}
\label{utilization_table}
\end{table*}

The RFSoC resources used to implement several SG-DMA configurations are given in Table \ref{utilization_table}, which were obtained by synthesizing a Vivado block diagram containing only the MPSoC, dual-channel SG-DMA, AXI Interconnect, AXI SmartConnect, and Processor System Reset IP blocks. These IP blocks represent the minimum set required for SG-DMA engine. The throughput results of this work characterize an AXI data-bus bit-width of 256. The SG-DMA engine throughput of other data bus widths is not reported in this paper. A system that leverages The AXIS channels of the SG-DMA engine are normally connected to a streaming interface in PL, but are left floating in this analysis to evaluate SG-DMA resource utilization. The table reports LUT, LUTRAM, flip-flop, and BRAM utilization as a function of the width of the MM2S and S2MM SG-DMA's data busses. The BRAM blocks are utilized by the AXI DataMover within the SG-DMA engine to create a FIFO buffer. The percentage of resources utilized is given with respect to the resources available on the ZCU111, and will vary depending on the device architecture. As reflected in the table, the base engine requires a fixed number of resources, e.g., on order of 10K LUTs, with resource utilization increasing nearly linearly with increasing bus width. BRAM utilization is an exception, where a maximum of 306 kb is reached with a bus width of 256 bits. 

\subsection{Customized State Machine (CSM)} \label{sec:StateMachine}

\begin{figure}
   \centering
   \includegraphics[width=0.48\textwidth,keepaspectratio=true]{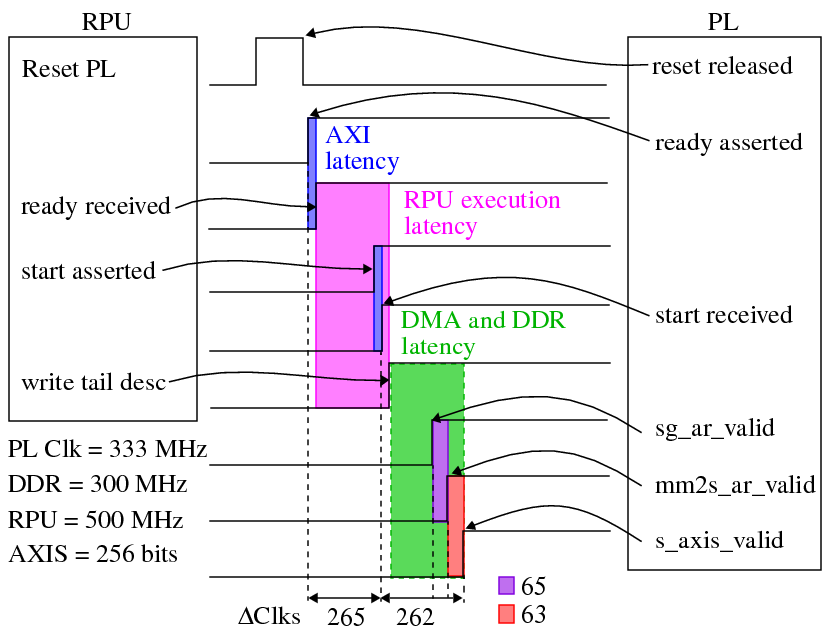}
   \caption{Latency components of the MM2S channel of scatter gather DMA associated with the first iteration. PL side clock frequency is 333 MHz, DMA engine configured with a bus width of 256-bits.}
   \label{PLTimingDiagram}
\end{figure}

The latency and throughput characteristics of the TIQC system architecture is measured using a customized state machine (CSM). An illustrative example showing the types of timing information that can be measured using the CSM is shown in Fig. \ref{PLTimingDiagram}. In this example, we show the interaction of the RPU with the SG-DMA and CSM components in the PL, and the corresponding counter values recorded within the CSM. The counter values reflect the latency of two components of the data transfer operation, namely, 1) execution time of the RPU and interactions across the GPIO interface associated with starting the SM and 2) latency of the SG-DMA measured between the start event and the arrival of the first word on the streaming interface in the PL. The sequence of operations are shown from top-to-bottom, while time (measured in clock cycles) is shown along the horizontal axis. The PL clock frequency in this example is 333 MHz while the SG-DMA bus width is configured to be 256-bits.

The details of this data transfer operation are given as follows. The RPU issues a reset to the SM through the GPIO as the initial step in the sequence. The first of two counters begins to increment after the release of reset. The RPU blocks and waits for the PL to assert a GPIO ready signal. Once received, the RPU asserts a GPIO start signal. The PL blocks (and increments the first counter) until the start signal is received. Once received, the first counter is stopped and the second counter is started. The reported 265 clock cycles accounts for the delays of the GPIO signal transmissions and RPU execution time. The PL then blocks and waits for the first word to be delivered through the streaming interface. This second time interval is broken down further by routing several AXI signals, namely \textit{sg\_ar\_valid}, \textit{mm2s\_ar\_valid}, and \textit{s\_axis} from the MM2S SG-DMA bus to the SM. These signals reflect the instant in time when the SG-DMA issues the BD address to DDR, the time instance when the SG-DMA engine issues the buffer address to DDR and when PL is notified that the first word from the BD buffer is available on the streaming interface, respectively. 

The reported values are the mean counter values associated with this RPU-bus-PL transaction. We report the mean, minimum, and maximum values converted to nanoseconds in the reminder of this paper. For example, the second counter value of 262 is equivalent to 787 ns of latency using a PL frequency of 333 MHz. A detailed description and ASMD diagram of the CSM is given in the Appendix.

\subsection{Experiment Overview} \label{sec:ExperimentOverview}

The objective of our analysis is to determine the performance characteristics, i.e., the average, best, and worst case data transfer characteristics, of a multi-processor architecture consisting of APU, RPU, and PL components. The partitioning of the tasks, namely, gate sequence management, BD ring creation, control, and execution, across these three components, enables the system to leverage the inherent parallelism that exists in the RFSoC platform. In our analysis, we report on the performance characteristics and elaborate on the speed-limiting components as they relate to the timing constraints of a TIQC system.

The interaction between the RPU and PL components, i.e., the SG-DMA engine and CSM, defines the inner loop of the TIQC system, and therefore possesses the tightest timing constraints. Our analysis of this subsystem explores parameters including the size of the BD ring, i.e., number of BDs and the size of the buffer regions associated with each BD (payload size). We also examine the impact of sequential versus random address locations of the BD payloads in PL DDR as a means of determining whether the DDR memory controller provides any look-ahead optimizations.

In a second subsection, we focus on analyzing the performance of the BD ring configuration process. The mean setup time and variability, i.e., minimum and maximum, are reported for different ring configurations. We compare the performance of the BD ring creation process when executed on the APU with the time taken when the same task is executed on the RPU, while accounting for the synchronization and communication latencies that are present when the APU creates the rings and hands-off to the RPU to schedule execution with the SG-DMA and CSM engines.

\begin{figure*}
   \centering
   \includegraphics[width=0.9\textwidth,keepaspectratio=true]{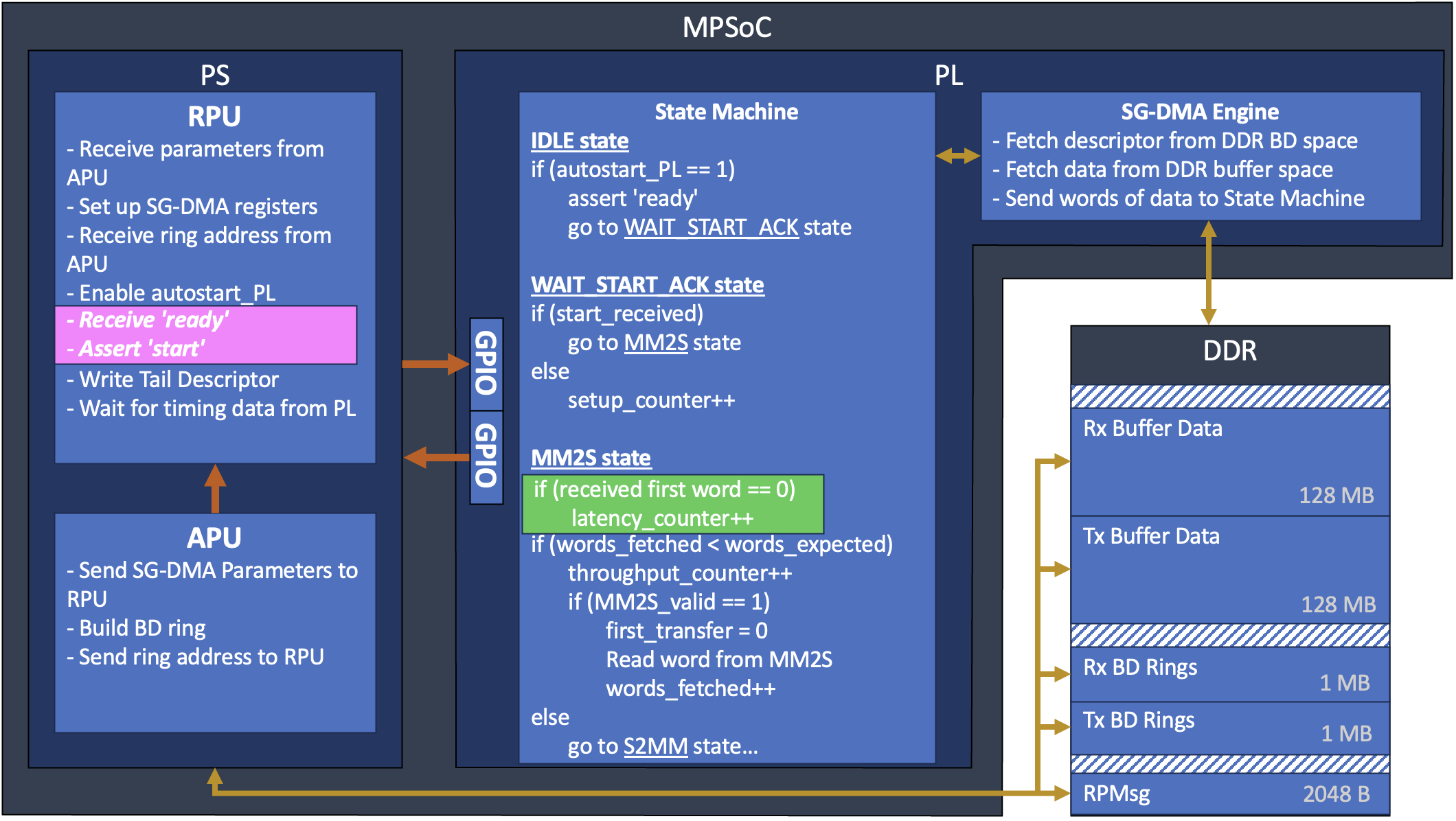}
   \caption{Multiprocessor SGDMA setup and state machine processes. The green and pink regions represent the timing regions in Fig \ref{PLTimingDiagram}. The pink timing region can be wrapped around any code running on APU or RPU to measure PS side execution time.}
   \label{SGDMA_FlowChart}
\end{figure*}

A block diagram illustrating the tasks and communication channels utilized in our experiments is shown in Fig. \ref{SGDMA_FlowChart}. The PL side remains the same across all experiments and consists of a CSM and SG-DMA engine. The timing events shown on the right side of Fig. \ref{PLTimingDiagram} are associated with the IDLE, WAIT\_START\_ACK, and MM2S states in the CSM. The three counters in the CSM, \textit{setup\_cnter}, \textit{MM2S\_Lat\_cnter}, and \textit{MM2S\_ThrPut\_cnter} are used to measure the corresponding performance characteristics of the data transfer operations. Note that the latter two counters are not shown in the CSM diagram but are configured identically to the MM2S versions. All components in the PL run at the maximum allowable frequency of 333 MHz. The PL DDR memory interface controller is also implemented in the PL and runs at 300 MHz. 

The CSM is connected to the SG-DMA Engine component through a 256-bit wide streaming AXIS interface. The data received on this interface is timing-critical because it is used to drive the IQ modulation system responsible via high-speed DACs integrated into the PL fabric. RPMsg utilizes shared memory in the PL DDR to enable data to be transferred between the APU and RPU. Alternatively, the APU and RPU can use the GPIO registers to exchange data, and given the GPIO are implemented in the PL, they are also accessible by the CSM, providing a three-way communication and synchronization channel between APU, RPU, and the CSM. The GPIO register can also be used within the CSM to obtain clock-cycle accurate timing measurements of communication latency between APU and RPU. We investigate the trade-offs of using RPMsg versus GPIO for APU-RPU-CSM communication and synchronization in the following sections.

The left side of Fig. \ref{SGDMA_FlowChart} shows the tasks carried out by the RPU and APU in the PS under the two BD-ring configuration options described above. In either scenario, the APU determines the number of BDs and defines the contents of BD buffers. In the first scenario, the RPU receives BD ring configuration parameters from the APU, e.g., the number of BDs, the buffer addresses, and the sequence order, and executes code to create the rings. In the second scenario, the APU creates the ring and only passes the base address of the ring to the RPU. In either scenario, the process of creating the SG-DMA data structure involves the following steps:

\begin{itemize}
    \item Allocate memory space for BD ring and buffer data
    \item Write test data into the buffer data region
    \item Create BD ring linked list with buffer address and status information, e.g., first BD and last BD, etc.
    \item Perform sanity checks on ring, flush cache
    \item Configure SG-DMA control registers
    \item Write to SG-DMA tail descriptor register
\end{itemize}

The last step, which starts the DMA transfer operation, is always performed by the RPU in our experiments. The RPU is also tasked with all synchronization operations that occur with the CSM. Once the transfer operation is complete, the PL notifies the RPU and transfers the counter values through the GPIO register interface. The RPU forwards the counter values to the APU using RPMsg and then to a desktop for off-line analysis.

Note that carrying out DMA in scatter gather mode requires the DMA engine to access DDR three times, once to fetch the BD, a second time to fetch the buffer data, and a third time to write an update the BD status register. Moreover, unlike the buffers themselves which may be allocated at sequential locations in the DDR, the BDs are always stored in a separate memory location. Therefore, SG-DMA possess additional overhead over simple mode that comes with a performance penalty. However, SG mode offers several advantages, including the ability to quickly construct customized gate sequences that leverage pre-computed gate sequences distributed across the DDR memory space.

Moreover, the utilization of DDR itself, in contrast to smaller embedded memory resources, e.g., PL BRAM, enables additional scalability and flexibility in the TIQC control system. The proposed system can easily leverage improvements to DDR transfer speed and capacity as a means of improving the throughput of the control system as well as the number and variety of gate sequences that the control system has access to. 

\subsection{Memory Map of the Experimental System}
\label{sec:ExperimentMemMap}

The PS and PL DDRs are utilized by the APU, RPU, and SG-DMA engine, with the APU running Linux on top of a virtual, paged memory system, while RPU and SG-DMA access physical memory directly, with the former running a bare metal application. As indicated, RPMsg also uses a shared DDR memory region for data exchange and synchronization. 

The organization of the DDR memories utilized in our experiments is shown in Fig. \ref{SGDMA_FlowChart}. The physical memory accessed by RPU and SG-DMA is excluded from the paged, virtual memory system of the APU by adding a reserved memory region to the Linux device tree in the 4 GB aperture at the address range between 0x80000000 and 0x9FFFFFFF. We construct BD rings and define BD buffers in this 512 MB region. Given the aperture is located within the 32-bit address space of the RPU, this strategy enables performance comparisons related to BD ring creation to be made between the APU and RPU. A more attractive arrangement in which the PL DDR is memory-mapped above the PS DDR would allow a TIQC system to fully utilize the 8 GB of DDR, but would restrict updates to the PL DDR to the APU.

The lower 2048 bytes of the DDR aperture is used for RPMsg communication between APU and RPU. The next 4 MB is allocated for BD rings, which allows rings of size $2^{16}$ BDs. The region above the BD rings of size 256 MB is allocated as BD buffer space. 

\section{Experimental Results}

Scatter gather mode provides a great deal of flexibility in creating a complex sequence of data transfer operations. The user controls several BD ring parameters including the number, size, and location of distinct buffer spaces. The performance characteristics of SG-DMA depend on the BD ring parameters and are the focus of our analysis in the next section. 

\begin{table}[]
    \centering
    \begin{tabular}{|l|l|l|l|}
    \hline
    Parameter & Bytes per BD & BDs per ring & Ring cycles \\ \hline
    Tested Range     & $2^{5}$ - $2^{13}$   & $2^{1}$ - $2^{13}$   & $2^{1}$ - $2^{13}$  \\ \hline
    \end{tabular}
    \caption{Range of tested SG-DMA ring parameters}
    \label{parameter_table}
\end{table}

In a separate set of experiments, we compare the performance of BD ring creation process on the APU and RPU across the range of BD parameters shown in Table \ref{parameter_table}, and evaluate the performance trade-offs between using RPMsg and GPIO as an information exchange and synchronization mechanism. Throughout all experiments, trials were run to evaluate the effect of repeatedly cycling through a single BD ring, a mode referred to as \textit{cyclic mode} in DMA literature. However, the performance characteristics of cyclic mode were observed to be identical to those measured using a ring with more BDs, and are therefore omitted. 

\subsection{SG-DMA Performance Analysis}

Latency is defined in our experiments as the interval of time associated with the writing of the tail descriptor and the arrival of the first word on the AXI MM2S streaming interface in the CSM. An example is shown in Fig. \ref{PLTimingDiagram}, where the second interval labeled '262' (787 ns) between events \textit{start received} and \textit{s\_axis\_valid} represents latency. Multiple repeated experiments are performed to obtain the median, minimum, and maximum values of latency, which are reported in units of nanoseconds in the following. Although the preceding synchronization events, i.e., events that occur in the region labeled '265', are also experienced in a typical implementation, we exclude them in our analysis. Similarly, the median, minimum, and maximum values of throughput are reported in MB/sec, and, in our analysis, we exclude the latency of the first event in their calculation. We use nonparametric statistical metrics, i.e., median, minimum and maximum, in contrast to the mean and standard deviation, because the measurements rarely follow a Gaussian distribution.

\begin{figure}
   \centering
   \includegraphics[width=3.7in,keepaspectratio=true]
    {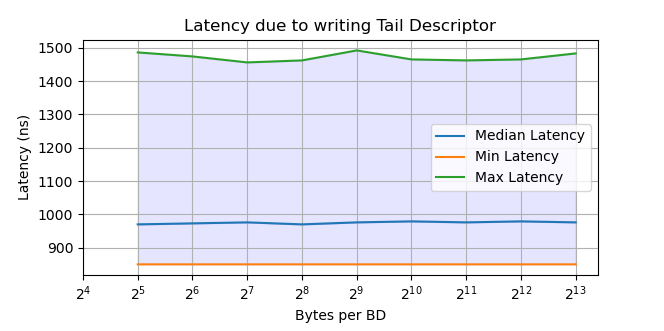}
   \caption{RPU-to-SG-DMA latency as a function of the number of bytes per BD.}
   \label{Latency_Variation}
\end{figure}

The results of the RPU-to-SG-DMA latency experiment are shown in Fig. \ref{Latency_Variation}. The x-axis plots the number of bytes per BD against latency on the y-axis in nanoseconds (ns). The results across all BD ring sizes are superimposed, and are indistinguishable, illustrating that latency is independent of the number of BDs, as expected. The shaded region shows the variation in the measurements, with a minimum latency of 849 ns up to a maximum latency of 1516 ns. Although the RPU architecture is built with features designed to minimize latency, it still follows the standard fetch-decode-execute-writeback sequence to perform operations and is therefore subject to typical microprocessor stall events related to access contention and refresh cycles in the DDR, cache misses, and AXI bus contention. For example, the DDR4 device used on the ZCU111 (MT40A512M16JY-075E) specifies a refresh row stall is required every 7.8\mics\!. We used Xilinx ChipScope and observed DDR stalls occurring for approximately 210 ns in duration, and asynchronously with program execution.

\begin{figure}
   \centering
   \includegraphics[width=3.7in,keepaspectratio=true]
    {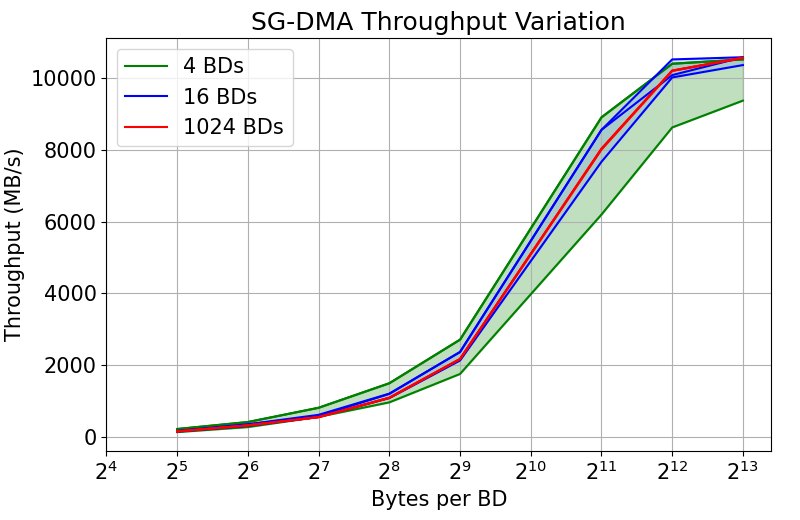}
   \caption{SG-DMA throughput as a function of the number of bytes per BD and BD ring size. The shaded regions highlight the variability and median throughput of the SG-DMA engine decreasing for larger BD rings.}
   \label{Thrput_variation}
\end{figure}

\begin{figure}
    \centering
    \includegraphics[width=3.3in,keepaspectratio=true]{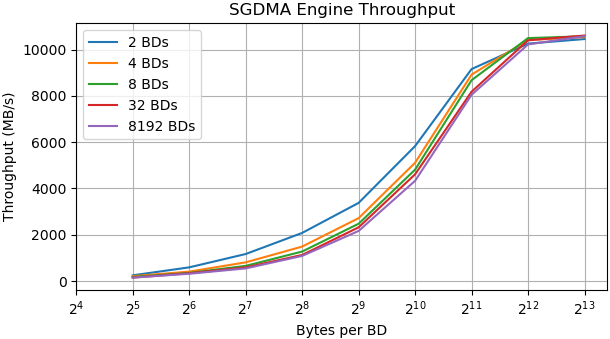}
    \caption{Throughput of the SG-DMA engine suffers a diminishing penalty as the number of BDs in the ring increases. Throughput increases linearly until saturating at the AXI data bus's maximum bandwidth of 10.6 GB/s.}
    \label{Throughput of SGDMA}
\end{figure}

The corresponding results for throughput are shown in Fig. \ref{Thrput_variation}, with bytes per BD plotted along the x-axis against throughput in MB/s on the y-axis. Unlike latency, throughput is impacted by the BD ring size. The graph superimposes the throughput results for BD rings of size 4, 16 and 1024. The variability in throughput is largest for the BD ring of size 4, and decreases linearly as the total amount of data transferred increases. Given the DDR stalls occur at regular time intervals, the impact that they have on the throughput associated with larger rings is amortized over the longer data transfer time interval, reducing their impact on variability. Moreover, the negative performance impact of thrashing between DDR memory locations corresponding to the BD ring and buffer addresses also becomes a smaller fraction of the overall transfer time for larger transfer sizes. The median throughput results for all tested BD ring sizes and buffer sizes is shown in Fig. \ref{Throughput of SGDMA}. The number of cycles has the same effect on throughput as the number of BDs, and buffer data can be stored discontiguously in memory without degrading throughput. 

To meet the lower-bounded shortest gate time of $\approx$1\mics, the throughput of the SG-DMA engine must be $\geq$ 32 MB/s. In the proposed TIQC system, the SG-DMA engine must transfer at least 32 bytes per BD to transfer a complete gate sequence. The worst case throughput for SG-DMA transfers of $\geq$ 32 bytes per BD measured in our experiments is 125 MB/s.

\subsection{BD Ring Creation Performance Analysis}

The BD ring creation process described earlier can be executed by either the APU or RPU, and consists largely of instantiating a SG-DMA data structure for managing BD rings and maintaining status. The SG-DMA IP block defines a suite of registers that Linux or bare metal API functions utilize for control and status operations. Both the APU and RPU access the physical memory address space of these registers directly. The C code running on the APU accomplishes the task of by-passing the virtual memory system by opening, reading and writing to the device in /dev/mem under Linux. The C code associated with the BD ring creation process is otherwise identical on both the APU and RPU. 

In order to provide an apples-to-apples performance comparison, the CSM is used to measure the $\Delta$t associated with the creation of the BD ring in both cases. From the timing diagram shown in Fig. \ref{PLTimingDiagram}, the counter which is sandwiched between GPIO signals \textit{Reset PL} and \textit{start asserted} can be used to measure execution time between any two points encapsulated by the C code which writes these two GPIO signals. The entire sequence of steps outlined in Section \ref{sec:ExperimentOverview} are the target of our timing operation. Note that the C code which writes to the buffer spaces, e.g., with gate sequences, is not included in the BD ring creation process, and instead, is handled by separate routines before this sequence is executed.

\begin{figure}
    \centering
    \includegraphics[width=3.3in,keepaspectratio=true]{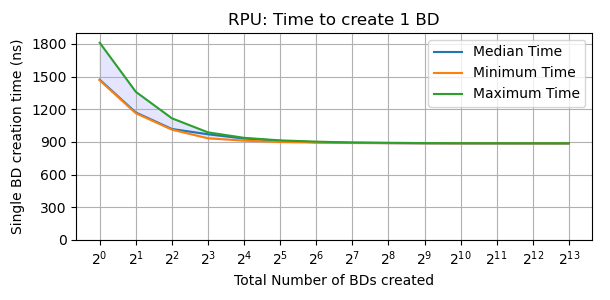}
    \caption{Time taken for RPU to create a single BD.}
    \label{RPU_1_BD}
\end{figure}

\begin{figure}
    \centering
    \includegraphics[width=3.3in,keepaspectratio=true]{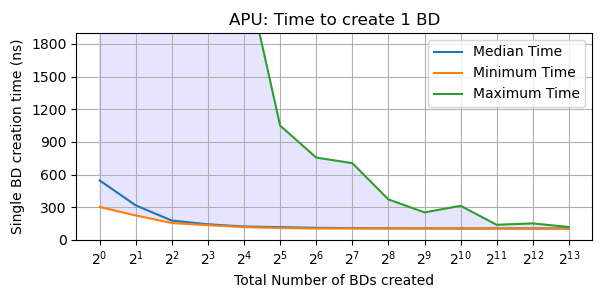}
    \caption{Time taken for APU to create a single BD.}
    \label{APU_1_BD_Scaled}
\end{figure}

\begin{figure}
    \centering
    \includegraphics[width=3.3in,keepaspectratio=true]{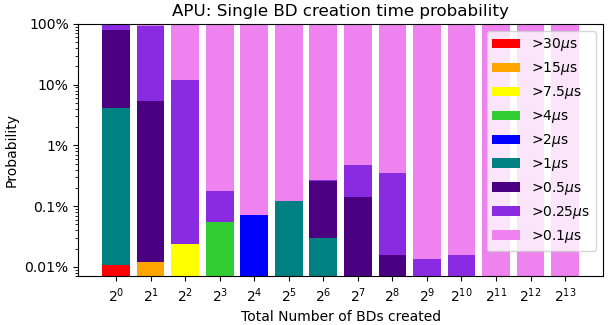}
    \caption{Probabilities for single BD creation time after a total of 100,000 data points. The observed best-case performance is a single BD taking an average of 0.1 to 0.25$\mu$s to be created, while the absolute worse case could take over 30$\mu$s. BD creation time was not observed to smoothly transition between the range of all reported values.}
    \label{APU_probabilities}
\end{figure}

The results of the performance comparison are shown in Figs. \ref{RPU_1_BD} and \ref{APU_1_BD_Scaled}. The execution time plotted on the y-axis is divided by the size of the BD ring, so the y-axis shows the average time to create a single BD in the ring. We observe that APU has a 5-9x speedup over RPU in the best-case scenario. However, APU suffers significantly longer stalls of up to 30\mics during execution of the user-space program when running a Linux 5.4.0 kernel created in PetaLinux 2020.2. These longer stalls in APU must be considered when attempting to meet timing requirements in a TIQC system. Fig. \ref{APU_probabilities} shows the ranges of APU execution time to create a single BD and their observed probabilities of occurrence.

\subsection{Inter-processor Data Transfer using RPMsg and GPIO}

The proposed TIQC system utilizes the high performance, more flexible APU for creating BD rings of gate sequences and for determining which gate sequences to run, and leverages the low execution time latency and variation of the RPU for control of the SG-DMA engine ring scheduling operations. This type of architectural arrangement requires the transfer of the selected BD ring address from the APU to the RPU. The two primary communication mechanisms by which APU can relay information to RPU is through RPMsg and GPIO. 
RPMsg defines an API for the exchange of data between APU and RPU. A 2048-byte shared memory region of DDR is reserved in the Linux device tree and initialized for RPMsg transfers. Receive (Rx) and transmit (Tx) buffer regions are defined by the API within this shared memory region for data transfer operations, as well as a mechanism for communicating status and executing handshake control operations for inter-processor synchronization and communication.

The BD ring address is transferred from APU to RPU by having the APU write the BD ring address into the Tx buffer then incrementing the Tx handshake flag. RPU checks the value of the flag in a busy wait loop and reads the address if its local copy is less than the shared value. Similarly, RPU increments a second Rx flag to acknowledge and complete the transfer. RPMsg also supports inter-processor interrupts for data transfers as an alternative to the busy-wait method described here \cite{irtija:2023}. The GPIO address transfer method uses a similar 2-way handshake method but implements the flags in PL-side hardware registers. 
Although both RPMsg and GPIO use the AXI interconnect to communicate, RPMsg additionally must contend for access to the DDR and is subject to stalls due to DDR refresh operations. Another salient benefit of GPIO is the potential of enabling three-way communication between APU, RPU and PL, given the registers are actually implemented in the PL.

\section{Application to Quantum Computing}
\label{sec:discussion}

The objective of this research was to develop and test Scatter Gather DMA for its ability to satisfy communication requirements for the control system of a trapped-ion quantum computer.  While SG-DMA is slower than simple-mode DMA, it has advantages for higher level sequencing of data streams where segments of the underlying data are reused. This applies to both the case where the segments addressed by BDs are small but frequently reused or large and only occasionally reused. These conditions are particularly applicable to quantum computing where electronic waveforms for quantum gates or shuttling ions may require large amounts of data but are frequently repeated, with only small changes needed to recalibrate them. Recalibrating gates using SG-DMA can be acheived efficiently by updating the data in the buffer which carries through to any gates that reference that BD. 

This type of scheduling offers more flexibility when large amounts of data are required, and overall cuts down on data communication requirements in the control system.  In this work we measured the dependence of throughput on BD size and found that SG-DMA can match the performance of simple-mode DMA when buffer sizes exceed $2^{12}$ bytes, showing that SG-DMA is a useful way to coordinate complicated but recycled buffer sequences (e.g. gate sequences or ion shuttling waveforms).  This framework is not just useful for trapped-ion quantum computing control systems, but can be applied to other quantum technologies like reconfiguring neutral atom arrays or performing composite pulse sequences in a superconducting quantum computer.

\section{Conclusion}
\label{sec:conclusion}

An experimental evaluation of latency and throughput of SG-DMA, implemented on a Xilinx RFSoC, the ZCU111, is carried out in this work, and the benefits and limitations described as they relate to the use of this device and architecture in TIQC control systems. Our analysis of overall system performance considers communication between elements of a multi-processing system, including the APU, RPU, and PL, using different workloads for the SG-DMA engine. Our conclusions are summarized as follows. 

\begin{itemize}
    \item The latency of the RPU to initiate SG-DMA engine operations can be characterized as having low variability and does not depend on the parameters of the transfer.
    \item The SG-DMA engine is most efficient when transferring large buffers regions, and has diminishing levels of variability in performance when processing large BD rings. The SG-DMA engine, though implemented in PL, does not have a fixed level of performance due to AXI bus and DDR contention.
    \item Repeatedly cycling through a single BD ring has the same effect on performance as increasing the number of BDs. 
    \item The nature of SG-DMA to store both BD rings and data buffers in DDR provides high levels of flexibility at the cost of reduced performance due to the additional memory accesses required to fetch BDs.
    \item With respect to application of the proposed architecture to TIQC systems, the SG-DMA engine's measured throughput is sufficient to meet gate sequence transfer requirements, and is in fact much better than the $\approx$1\mics lower-bound target speed.
    \item GPIO registers can be utilized as an alternative to RPMsg for low throughput inter-processor communication and data transfer, and provides a potential 3-way communication mechanism between APU, RPU and custom state machines in the PL. 
    \item The APU provides a 5-9x median execution time speedup over RPU when creating BD rings. However, APU has significantly slower worst-case performance for small numbers of BDs than RPU primarily due to Linux interrupt handling. The trade-off of high performance or low variability motivates the usage of APU and RPU for tasks that best suit their respective strengths.
    \item APU is better equipped for complex calculations by virtue of its L2 cache, 64-bit architecture and larger number of cores, while RPU is more appropriate for timing-critical portions of the control system such as reading/writing data to and from the PL.
\end{itemize}

Although the analysis carried out in this work focuses on the application of the RFSoC to TIQC system architectures, the results are relevant for other QC control systems and alternative RFSoC and MPSoCs. The proposed multi-processor system architecture leverages parallelism across the APU, RPU and PL state machine computational units to effectively improve the performance, flexibility and scalability of TIQC control systems. 


\section*{Acknowledgment}
This work is supported by a collaboration between the US DOE and other Agencies.  This material is based upon work supported by Quantum Systems through Entangled Science and Engineering through National Science Foundation Quantum Leap Challenge Institutes under Grant OMA-2016244. This material is also based upon work supported by the U.S. Department of Energy, Office of Science, National Quantum Information Science Research Centers, Quantum Systems Accelerator, and by the U.S. Department of Energy, Office of Science, Office of Advanced Scientific Computing Research Quantum Testbed Program.

Sandia National Laboratories is a multimission laboratory managed and operated by National Technology \& Engineering Solutions of Sandia, LLC, a wholly owned subsidiary of Honeywell International Inc., for the U.S. Department of Energy’s National Nuclear Security Administration under contract DE-NA0003525.  This paper describes objective technical results and analysis. Any subjective views or opinions that might be expressed in the paper do not necessarily represent the views of the U.S. Department of Energy or the United States Government.

\section*{APPENDIX}
\label{sec:appendix}

\subsection{CSM ASM Diagram} \label{sec:ASMD}

\begin{figure}
    \centering
    \includegraphics[width=0.35\textwidth,keepaspectratio=true]{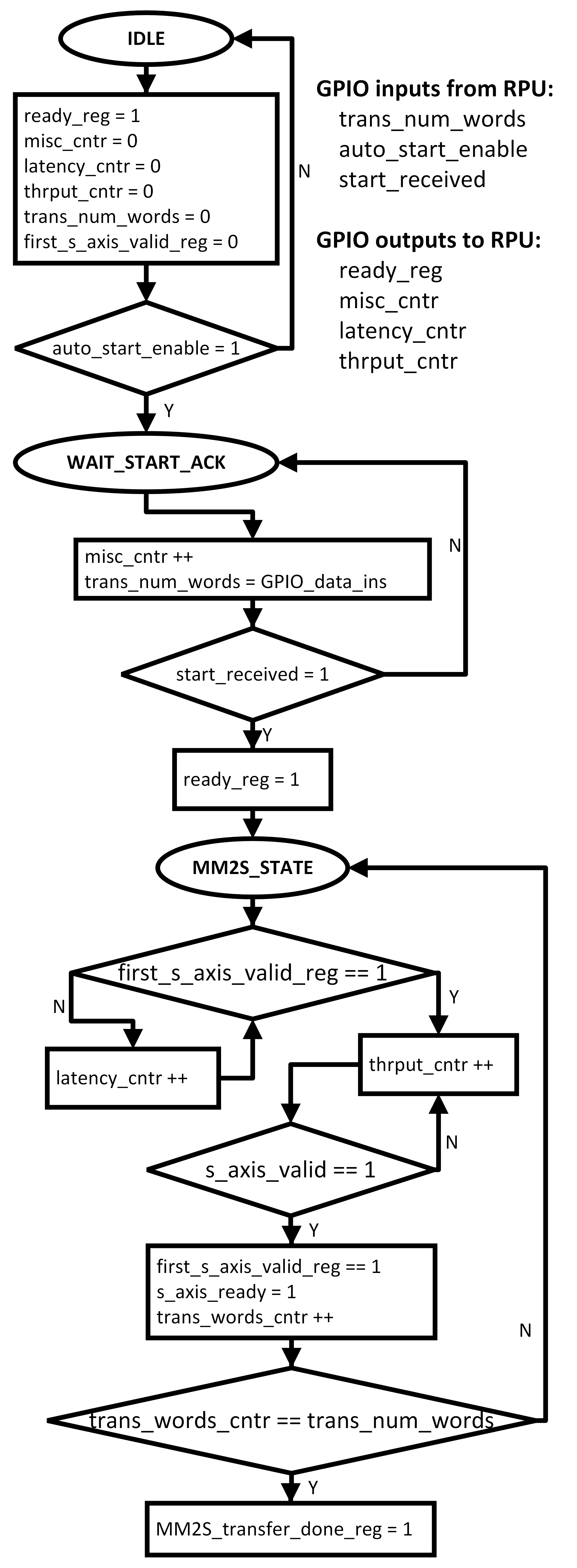}
    \caption{ASMD diagram of the PL state machine that is used as the timing engine and as the primary interface to the SG-SMA engine.}
    \label{ASMD_SGDMA}
\end{figure}

To measure the performance of the SG-DMA engine, a PL timing state machine was designed that can send/receive a single 256 bit transfer on every clock cycle. With the PL synthesized using the ZCU111's maximum PL clock speed of 333MHz, this sets the maximum throughput of the system at 256 bits every 3.003 ns, or 10.6 GB/s. To measure the performance of the processors, an additional PL timing state machine was implemented that can be started and stopped by the processors, thus measuring the execution time of any PS-side operation. Details related to the PS-side timing operations are presented in Fig. \ref{PLTimingDiagram}. An algorithmic state machine diagram of the PL customized state machine is presented in Fig. \ref{ASMD_SGDMA}, and described as follows.
\begin{enumerate}
    \item \textbf{IDLE}: Initialize all throughput and latency counters to 0, reset BRAM addresses for DMA transfer storage, reset flags for first word received and transfer completion. Busy wait for the RPU start signal, transferred through a GPIO register.
    \item \textbf{WAIT\_START\_ACK}: An intermediate state containing a counter and registers for a handshake operation between PL and PS that is used for measuring PS execution time.
    \item \textbf{MM2S\_STATE}: The state responsible for measuring the latency and throughput of the SG-DMA engine. The MM2S latency counter is used to measure the time interval between the writing of the tail descriptor by the RPU and the delivery of the first word through the streaming interface. The throughput counter is used to measure the time interval required to transfer the entire payload through the streaming interface, excluding the latency time interval. 
    \item \textbf{WAIT\_DONE\_GET\_CNT\_VALS}: Load counter values into GPIO registers for RPU to obtain. Wait for RPU to acknowledge it received all counter values.
\end{enumerate}

The state machine asserts its ready signal, begins incrementing the PS execution timer, and the PS then asserts its own start signal. The state machine then moves to the MM2S state where it begins incrementing the latency counter. Once the RPU receives the PL acknowledgement, it writes the tail descriptor into the SG-DMA engine, starting the MM2S transfer. The RPU then goes into standby while the PL completes the MM2S transfer. Once the SG-DMA engine raises MM2S\_valid, the state machine stops counting latency cycles, and starts counting throughput cycles. The state machine reads the 32 bytes from the SG-DMA engine for each MM2S\_valid signal assertion. The PL continuously reads from the SG-DMA engine until it receives the entire payload. In parallel, the SG-DMA engine is processing and updating the BD ring and fetching and sending 32 bytes of the buffer data from DDR to the state machine. Once the MM2S operations are completed, the state machine sends the counter values to the PS side for post-processing.

\clearpage

\nocite{*}

\bibliography{bibliography}

\end{document}